\pdfoutput=1
\documentclass[aps,prl,groupedaddress,superscriptaddress,showpacs,reprint,nofootinbib,nolongbibliography]{revtex4-2}
\usepackage{amsmath}
\usepackage{siunitx}
\usepackage{hyperref}
\usepackage{graphicx}
\usepackage{stackengine}
\usepackage{color}
\usepackage{subfigure}

\usepackage{helvet}
\usepackage{times}
\usepackage{enumitem}
\usepackage{svg}
\newcommand{\Jiuzhang}{\textit{Ji\v{u}zh\={a}ng\ }}

\begin{document}
\title{Gaussian Boson Sampling with Pseudo-Photon-Number Resolving Detectors and \\ Quantum Computational Advantage}

	\author{Yu-Hao Deng}
	\author{Yi-Chao Gu}
	\author{Hua-Liang Liu}
	\author{Si-Qiu Gong}
	\author{Hao Su}
	\author{Zhi-Jiong Zhang}
	\author{Hao-Yang Tang}
	\author{Meng-Hao Jia}
	\author{Jia-Min Xu}
	\author{Ming-Cheng Chen}
	\author{Jian Qin}
	\author{Li-Chao Peng}
	\author{Jiarong Yan}
	\author{Yi Hu}
	\affiliation{Hefei National Laboratory for Physical Sciences at Microscale and School of Physical Sciences, University of Science and Technology of China, Hefei, Anhui, 230026, China}
	\affiliation{CAS Centre for Excellence and Synergetic Innovation Centre in Quantum Information and Quantum Physics, University of Science and Technology of China, Shanghai, 201315, China}
	\affiliation{Hefei National Laboratory, University of Science and Technology of China, Hefei 230088, China}

	\author{Jia Huang}
	\author{Hao Li}
	
	\affiliation{State Key Laboratory of Functional Materials for Informatics, Shanghai Institute of Micro system and Information Technology (SIMIT), Chinese Academy of Sciences, 865 Changning Road, Shanghai, 200050, China}
	
	\author{Yuxuan Li}
	\author{Yaojian Chen}
	
	\affiliation{Department of Computer Science and Technology and Beijing National Research Center for Information Science and Technology, Tsinghua University, Beijing, China}
	
	\author{Xiao Jiang}
	
	\affiliation{Hefei National Laboratory for Physical Sciences at Microscale and School of Physical Sciences, University of Science and Technology of China, Hefei, Anhui, 230026, China}
	\affiliation{CAS Centre for Excellence and Synergetic Innovation Centre in Quantum Information and Quantum Physics, University of Science and Technology of China, Shanghai, 201315, China}
	\affiliation{Hefei National Laboratory, University of Science and Technology of China, Hefei 230088, China}
	
	\author{Lin Gan}
	\author{Guangwen Yang}
	
	\affiliation{Department of Computer Science and Technology and Beijing National Research Center for Information Science and Technology, Tsinghua University, Beijing, China}
	
	\author{Lixing You}
	
	\affiliation{State Key Laboratory of Functional Materials for Informatics, Shanghai Institute of Micro system and Information Technology (SIMIT), Chinese Academy of Sciences, 865 Changning Road, Shanghai, 200050, China}
	
	\author{Li Li}
	\author{Han-Sen Zhong}
	\author{Hui Wang}
	\author{Nai-Le Liu}
	
	\affiliation{Hefei National Laboratory for Physical Sciences at Microscale and School of Physical Sciences, University of Science and Technology of China, Hefei, Anhui, 230026, China}
	\affiliation{CAS Centre for Excellence and Synergetic Innovation Centre in Quantum Information and Quantum Physics, University of Science and Technology of China, Shanghai, 201315, China}
	\affiliation{Hefei National Laboratory, University of Science and Technology of China, Hefei 230088, China}

    \author{Jelmer J. Renema}
    
    \affiliation{Adaptive Quantum Optics Group, Mesa+ Institute for Nanotechnology, University of Twente, P.O. Box 217, 7500 AE Enschede, Netherlands}
	
	\author{Chao-Yang Lu}
	
	\affiliation{Hefei National Laboratory for Physical Sciences at Microscale and School of Physical Sciences, University of Science and Technology of China, Hefei, Anhui, 230026, China}
	\affiliation{CAS Centre for Excellence and Synergetic Innovation Centre in Quantum Information and Quantum Physics, University of Science and Technology of China, Shanghai, 201315, China}
	\affiliation{Hefei National Laboratory, University of Science and Technology of China, Hefei 230088, China}
	\affiliation{New Cornerstone Science Laboratory, Shenzhen 518054, China}
	
	\author{Jian-Wei Pan}
	
	\affiliation{Hefei National Laboratory for Physical Sciences at Microscale and School of Physical Sciences, University of Science and Technology of China, Hefei, Anhui, 230026, China}
	\affiliation{CAS Centre for Excellence and Synergetic Innovation Centre in Quantum Information and Quantum Physics, University of Science and Technology of China, Shanghai, 201315, China}
	\affiliation{Hefei National Laboratory, University of Science and Technology of China, Hefei 230088, China}

\begin{abstract}
We report new Gaussian boson sampling experiments with pseudo-photon-number-resolving detection, which register up to 255 photon-click events. We consider partial photon distinguishability and develop a more complete model for the characterization of the noisy Gaussian boson sampling. In the quantum computational advantage regime, we use Bayesian tests and correlation function analysis to validate the samples against all current classical spoofing mockups. Estimating with the best classical algorithms to date, generating a single ideal sample from the same distribution on the supercomputer \textit{Frontier} would take $\sim 600$ years using exact methods, whereas our quantum computer, \Jiuzhang 3.0, takes only \SI{1.27}{\us} to produce a sample. Generating the hardest sample from the experiment using an exact algorithm would take \textit{Frontier} $\sim \num{3.1e10}$  years.
 \end{abstract}
 \maketitle
 
 Quantum computational advantage (QCA) \cite{bernsteinQuantumComplexityTheory1993, preskillQuantumComputingEntanglement2012, harrowQuantumComputationalSupremacy2017,lundQuantumSamplingProblems2017,hangleiterComputationalAdvantageQuantum2022} marks an important milestone in the development of quantum computers. By solving certain quantum sampling problems \cite{Aaronson2011,boixo2018characterizing} that are intractable for classical supercomputers, QCA experiments \cite{aruteQuantumSupremacyUsing2019, Zhong2021,  zhongPhaseProgrammableGaussianBoson2021, wuStrongQuantumComputational2021, madsenQuantumComputationalAdvantage2022} have provided strong evidence for the long-anticipated quantum speed-up theoretically conceived $\sim 40$ years ago,
 Similar to Bell tests \cite{bellEinsteinPodolskyRosen1964} which were designed to refute the local hidden variable theories \cite{einsteinCanQuantumMechanicalDescription1935}, these QCA experiments offered evidence of the violations of the Extended Church-Turing thesis \cite{harrowQuantumComputationalSupremacy2017}.
 
Interestingly, these QCA experiments have motivated growing study of faster classical simulations. For boson sampling, these efforts can be divided into three approaches.
The first approach is reducing the classical simulation overhead of exactly simulating an ideal implementation of the Gaussian boson sampling protocol \cite{kaposiPolynomialSpeedupTorontonian2021, 
quesadaQuadraticSpeedUpSimulating2022,
bulmerBoundaryQuantumAdvantage2022}. In this case, one expects that the exponential gap between the classical and the quantum persists, but it is possible to bring down the overhead. The leading result is Ref. \cite{bulmerBoundaryQuantumAdvantage2022} which uses photon collisions to achieve better exact classical simulation that could reduce the computational overhead by orders of magnitudes lower than brute-force algorithms.
The second approach is using imperfections in the experimental setup \cite{qiRegimesClassicalSimulability2020,
renemaSimulabilityPartiallyDistinguishable2020,shiEffectPartialDistinguishability2022}, such as photon loss and partial photon distinguishability, to come to a faster simulation of the experimental data produced by an imperfect experiment. In this case, one can hope for more efficient classical simulation if the level of imperfections or experimental noise become strong enough. The final approach is spoofed distributions \cite{bulmerBoundaryQuantumAdvantage2022,villalongaEfficientApproximationExperimental2022, ohClassicalSimulationBoson2022, ohSpoofingCrossEntropy2022}, i.e., distributions which are not constructed to be close in variational distance to the ideal boson sampling distribution, but which are designed to reproduce some statistical aspect of the experiment. In particular, the recently proposed more competitive squashed state mockup \cite{madsenQuantumComputationalAdvantage2022} exhibits better agreement with the ground truth than the thermal state mockup. Moreover, the treewidth sampler \cite{ohClassicalSimulationBoson2022}, designed to spoof experiments with restricted circuit-connectivity, and the Independent Pairs and Singles (IPS) sampler \cite{bulmerBoundaryQuantumAdvantage2022}, designed to emulate experiments of limited quantum interference, have not been reported to be ruled out in previous GBS QCA experiments.

In return, these new challenges from the classical counterparts motivate the development of higher-fidelity and larger-scale quantum computers, new methods for validation and characterization \cite{shchesnovich2021distinguishing,
delliosValidationTestsGBS2022,
seron2022efficient,
giordaniCertificationGaussianBoson2023}, as well as better modeling and understanding of the increasingly complex system \cite{
drummondSimulatingComplexNetworks2021,popovaCrackingQuantumAdvantage2021,
grierComplexityBipartiteGaussian2022,
limApproximatingOutcomeProbabilities2022, delliosSimulatingMacroscopicQuantum2022,
iosue2022page,
qiao2022entanglement,
liuComplexityGaussianBoson2023, ohClassicalSimulationAlgorithms2023}, which is a fundamental endeavor in its own right \cite{rahimi-keshariSufficientConditionsEfficient2016, deshpandeDynamicalPhaseTransitions2018,chabaud2023resources}. Indeed, only by such a continuous quantum-classical competition can the QCA milestone be progressively better established.

\begin{figure*}[!htp]
    \centering
    \includegraphics[width = 0.88\linewidth]{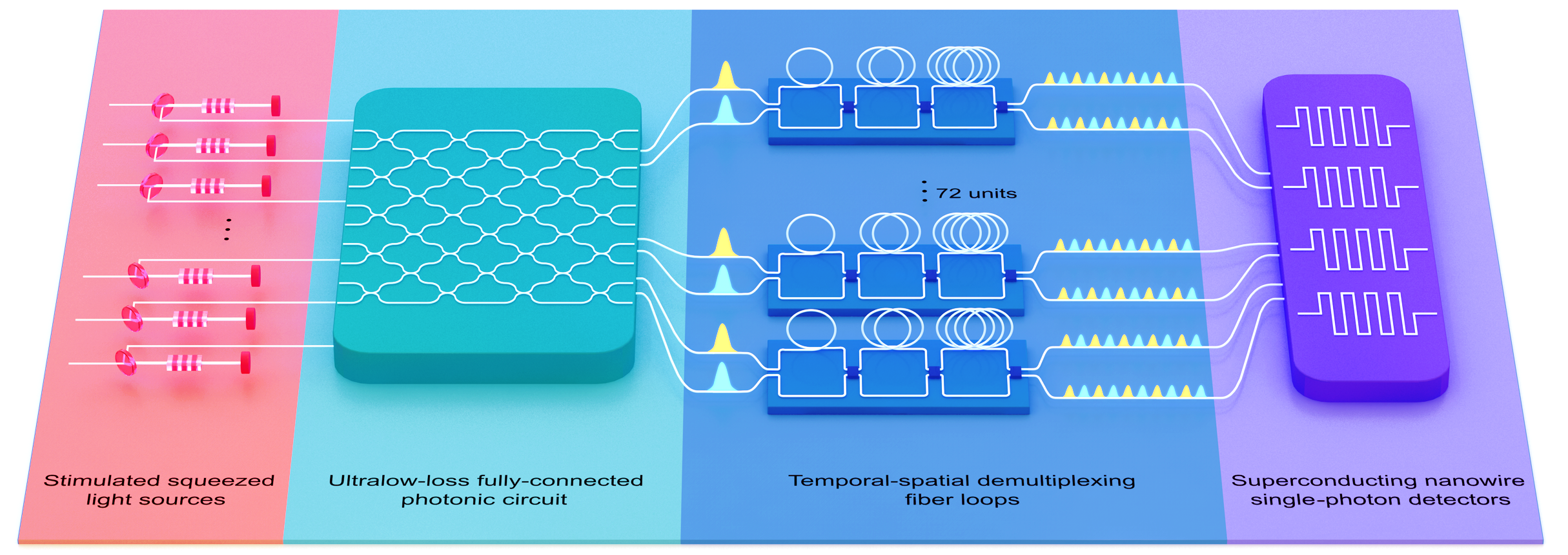}
    \caption{The experimental setup. 25 stimulated two-mode squeezed state photon sources are all phase-locked to each other and sent into a 144-mode ultralow-loss fully-connected optical interferometer. The photons go through 72 units of fiber loop setups for temporal-spatial demultiplexing and are detected by 144 superconducting nanowire single-photon detectors, which together constitute the pseudo-photon-number resolving detection scheme. Each fiber loop setup includes two input modes as represented by distinct colors. Photons from each mode are temporally demultiplexed by the fiber beam splitters and the delay lines into four time bins, and each time bin is furthermore split into two path bins at the last fiber beam splitter.
    The photons corresponding to each of the two input modes of the same fiber loop setup can be distinguished by their parity in time bin with a coincidence event analyzer (not shown).}
    \label{fig:1}
\end{figure*}

In this direction, we report in this \textit{Letter} a new, higher-efficiency Gaussian boson sampling (GBS) \cite{Hamilton2017, quesadaGaussianBosonSampling2018} machine with up to 255 photon clicks in the output using pseudo-photon-number-resolving detectors (PPNRD), which overcomes both the most powerful exact classical simulation algorithms and all known approximate algorithms and spoofing algorithms. The PPNRD scheme overcomes the shortcoming of threshold detection that it is incapable of resolving photon-number information, and significantly increases the exact sampling task’s computational complexity by more than six orders of magnitude.
We validate the samples against all the emerging classical mockups, particularly including the squashed state \cite{madsenQuantumComputationalAdvantage2022}, the treewidth sampler \cite{ohClassicalSimulationBoson2022} and the IPS sampler \cite{bulmerBoundaryQuantumAdvantage2022} mentioned above. A new model that includes the partial photon indistinguishability is used to characterize the system, and exhibits better agreement with the experiment than previous modelling methods. The computational complexity of this new GBS device, \Jiuzhang 3.0, is analyzed and a new QCA frontier is established.

The first GBS experiments \cite{Zhong2021, zhongPhaseProgrammableGaussianBoson2021} in the QCA regime used threshold detectors to register the samples. In those cases, there was a possibility of photon collision, that is, the photons can bunch at the outputs. New classical algorithms \cite{bulmerBoundaryQuantumAdvantage2022} could exploit photon collision to reduce the simulation overhead. A recent work \cite{madsenQuantumComputationalAdvantage2022} has reported time-bin-encoded GBS with photon-number resolved detection, in a fiber loop-based configuration similar to an earlier single-photon boson sampling experiment \cite{heTimeBinEncodedBosonSampling2017}. However, the relatively high photon loss in the fiber loops has allowed only three loops in the implementation, restricting the depth and universality of the interferometer. Additionally, the long recovery time of the transition-edge sensors could render it an unsuitable choice for high-repetition-rate experiments.

A schematic of our GBS experiment is shown in Fig. \ref{fig:1}. Transform-limited laser pulses double pass periodically poled potassium titanyl phosphate (PPKTP) crystals to create 25 pairs of two-mode squeezed state (TMSS) by a stimulated emission process \cite{zhongPhaseProgrammableGaussianBoson2021}. The TMSSs have an average coupling efficiency of 88.4\% and photon indistinguishability of 96.2\%, simultaneously. In the experiment, the laser power is tuned to generate different squeezing parameters ranging from 1.2 to 1.6. The TMSSs are then fed into an ultralow-loss three-dimensional interferometer with full connectivity among all 144 modes. The transmission rate of the interferometer is 97\% for each mode, and the average wave-packet overlap inside the interferometer is above 99.5\%. The whole set-up is actively phase locked within a precision of 15 nm.

For detection, we implement pseudo-photon number resolution using one-to-eight demultiplexing of the optical modes (see Fig. \ref{fig:1}).
We conduct detector tomography \cite{lundeenTomographyQuantumDetectors2009} of each mode to validate the pseudo-photon-number-resolved detection (PPNRD) scheme \cite{suppl}. As shown in Fig. S1(a), the probability of the nine photon-click number response of the PPNRD agrees well with the theoretical model. The positive-operator valued measures of the nine photon-click number measurement can be constructed from the data, which is shown in Fig. S1(c), together with its Wigner function and fidelity with the theoretical prediction. While we model the experiment with the PPNRD detection scheme in the following discussions, we address the question of how well our detection scheme emulates true photon number-resolving detection in \cite{suppl}.

\begin{figure*}[!htp]
    \centering
    \includegraphics[width = 0.85 \textwidth]{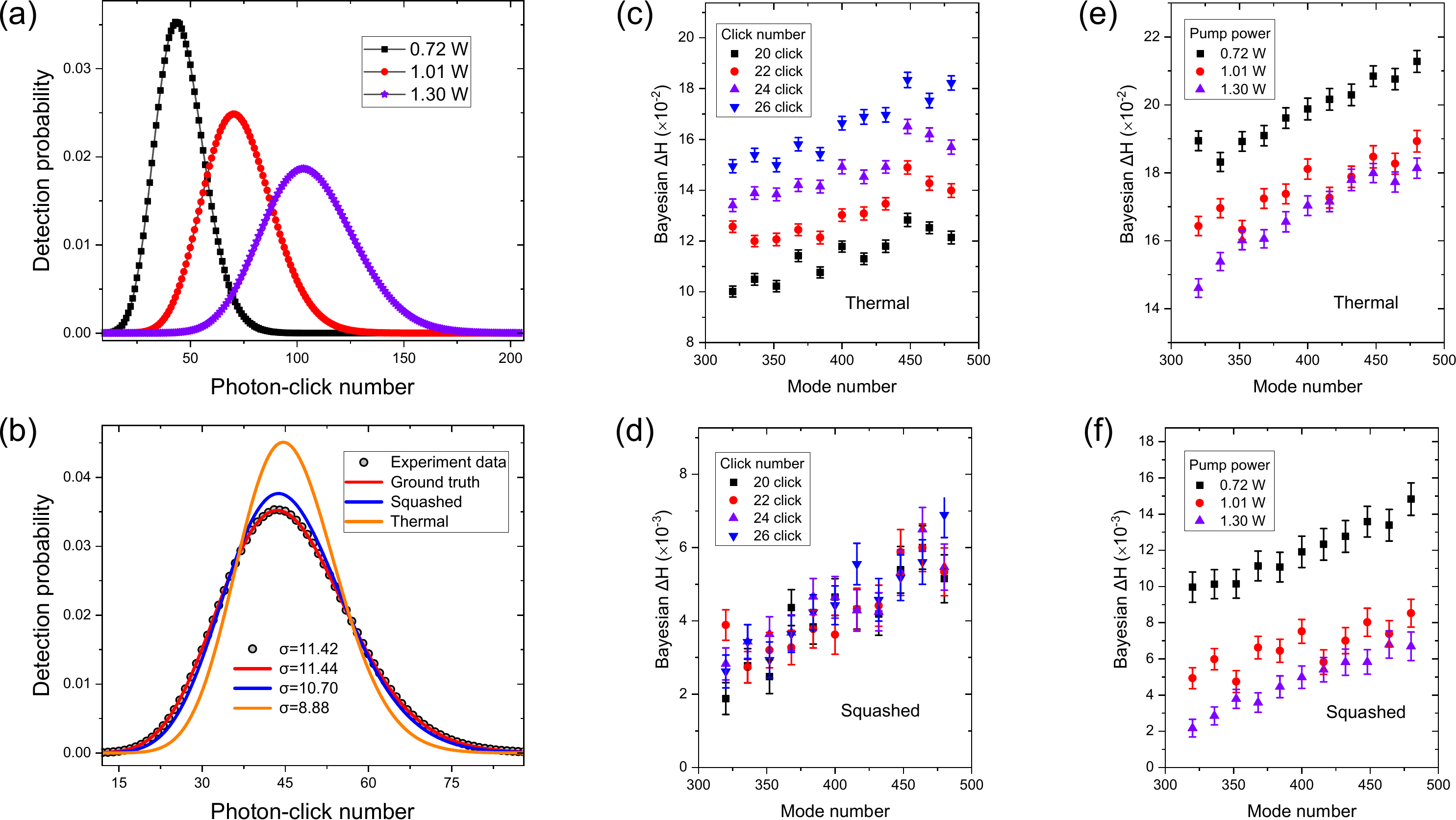}
    \caption{The experimental photon-click number distribution and the Bayesian validation results. (a) Photon-click number distribution of this work. Data from experiments of three pump laser power ranging from 0.72 W and 1.30 W are displayed. A maximum photon-click number of 129, 203, and 255 are registered for each of the experiments. (b) Photon-click number distribution of the experimental results, the ground-truth theory, the squashed state and the thermal state mockups for the lowest laser intensity configuration. The plot includes labels indicating the standard deviation of the photon-click number for each distribution. It is evident that the experimental distribution agrees with the ground truth best. The error bars are too small to be displayed. (c-d) Bayesian confidence of the ground-truth theory as a function of subsystem size, against the thermal state (c) and the squashed state (d) hypothesis in the high power (1.30 W) experiment. It can be observed that as the subsystem size grows, the Bayesian confidence is above zero and exhibits a clearly increasing trend, indicating an exclusion of the two mockups and a stronger Bayesian confidence for the full system. Bayesian results for photon click number changing from 20 to 26 are displayed in data points of different colors and show similar results. (e-f) Bayesian confidence of the ground-truth theory as a function of the subsystem size for experiments of the three distinct pump laser power. The results show the Bayesian results against the thermal state (e) and the squashed state (f) hypothesis, for a fixed photon-click number (25). For all pump laser powers, the Bayesian $\Delta H$ is above zero and exhibits a clearly increasing trend with subsystem size, implying stronger Bayesian confidence for the full system. In all plots error bars indicate standard error. The statistical fluctuation from photon-click number probability estimated with Monte-Carlo methods \cite{drummondSimulatingComplexNetworks2021} has been included in the error bars.
    For results of each mode number, the Bayesian score are obtained by averaging over an ensemble of randomly selected subsystems for unbiased benchmarking. 
}
    \label{fig:2}
\end{figure*}

\begin{figure*}[!htp]
    \centering
    \includegraphics[width = 0.8 \textwidth]{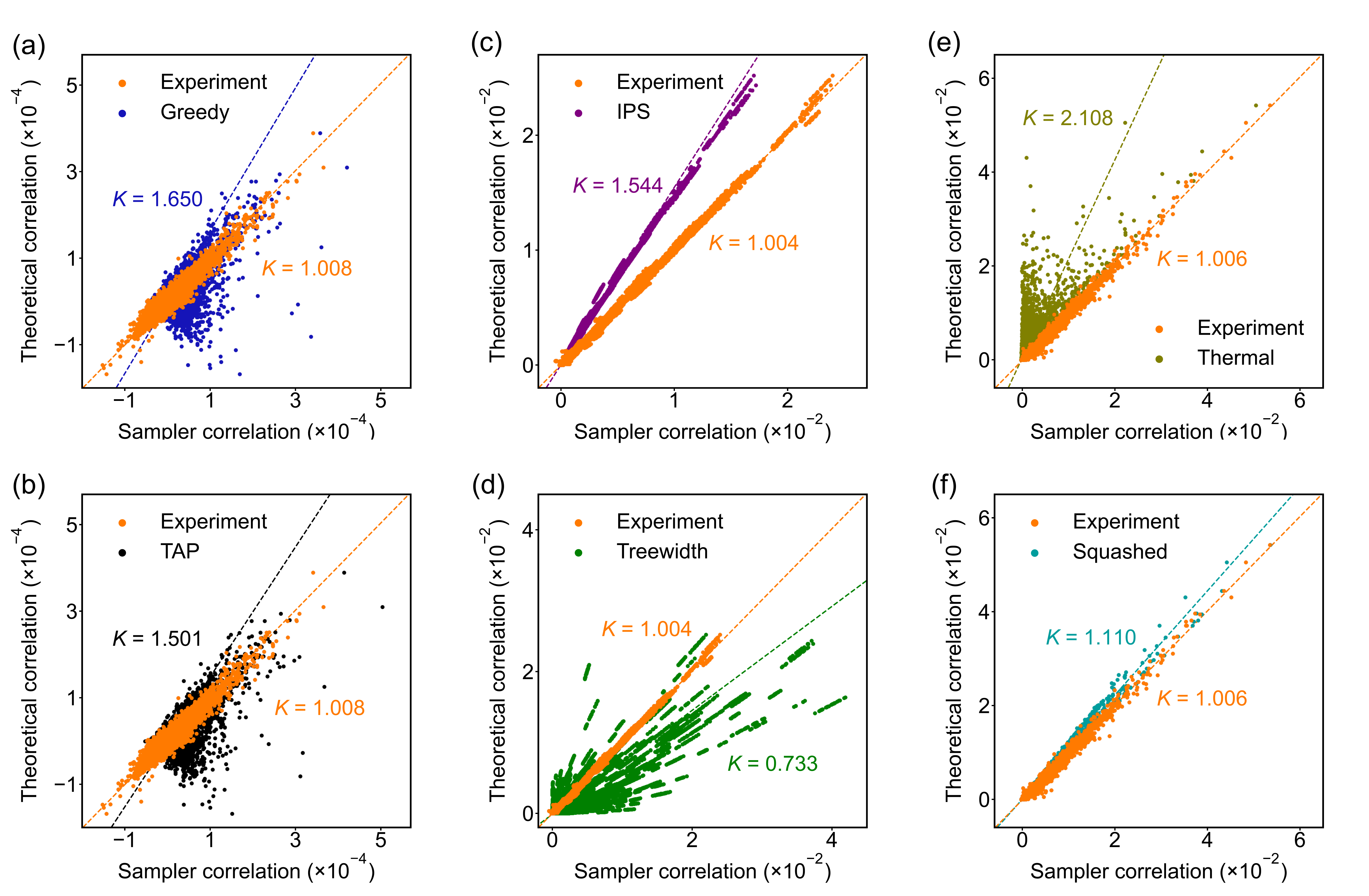}
    \caption{Correlation function analysis. The experiment clearly exhibits a better agreement with the ground truth than the mockups, with two-norm distance $D$ and slope $K$ (marked in colored dashed lines on each plot). (a-b) 3-order correlation function in scatter plot showing the experiment ($D=0.0054, K=1.008$), the greedy sampler (a) ($D=0.0062, K=1.650$), and the TAP sampler (b) ($D=0.0067, K=1.501$) in comparison versus the ground-truth theory. For (a) and (b), 10 million samples are used for estimation of the experimental and mockup’s correlation. All 3-order correlation functions of 144 out of 1152 modes (1 from each 1-to-8 fan-out modes) are displayed. (c-d) Scatter plot of the 2-order correlation function of the experiment ($D=0.045$, $K=1.004$), the IPS sampler (c) ($D=0.254$, $K=1.544$), and the treewidth sampler (d) ($D=0.739$, $K=0.733$), versus the ground-truth theory. In (c-d), all pairs of 1152 modes are displayed, and 10 million samples are used for estimation of the experimental correlation whereas exact theoretical value is used for the mockup samplers. (e-f) The 2-order correlation function of the experiment ($D=0.040, K=1.006$), the thermal state mockup (e) ($D=0.275, K=2.108$), and the squashed state mockup (f) ($D=0.052, K=1.110$) versus the ground truth theory, in coarse-grained 144 modes. In (e-f), 10 million samples are used for estimation of the experimental correlation and the mockup samplers' correlation. All data are from the highest power (1.30W) experiment.
}
    \label{fig:3}
\end{figure*}

The overall linear efficiency of the whole experimental set-up is 43\%, including the quantum light sources, transmission, and detection. This efficiency is much higher than that in ref. \cite{madsenQuantumComputationalAdvantage2022}, which was 33\% although involving only three loops. The photon-click number distribution under three laser powers is shown in Fig. \ref{fig:2}(a). The maximum photon-click number reaches 255, which is higher than all the previous GBS experiments.

We analyze the obtained GBS samples and validate them against known classical hypotheses in the QCA regime. The most powerful method to spoof the GBS is to use classical states which can maximally approximate the experiment's quantum light sources which, under photon loss, can be gradually degraded into squeezed thermal states \cite{qiRegimesClassicalSimulability2020}. Thus, in this work, in addition to the thermal state hypothesis which has been tested in \Jiuzhang 1.0 and 2.0, we consider the more competitive hypothesis, namely the squashed state \cite{madsenQuantumComputationalAdvantage2022, qiRegimesClassicalSimulability2020}. The squashed state is a classical mixture of coherent states and has vacuum fluctuations in one quadrature and larger fluctuation in the other.
It can be optimized to have the same mean photon number as the input lossy squeezed state \cite{ClassicalModelsAre2022}, which is the convention we take in this work. Note that there are other plausible hypotheses such as using coherent light and distinguishable photons, which are much easier to be ruled out. As shown in \cite{suppl}, the validation strength against these other hypotheses is typically 2-4 orders of magnitude larger than the squashed states. Therefore, we will focus on the discussion of the latter in the main text.

We start by comparing the experimental photon number distribution with the ground-truth model and the possible mockups in Fig. \ref{fig:2}(b). The GBS data set taken at \SI{0.72}{W} laser power (dot) well overlaps with the ground truth (red), while obviously deviates from the classical mockups based on the thermal state (yellow) and the squashed state (blue). These distributions can be quantitatively distinguished by their standard deviation as labelled on the plot, where the experiment exhibits the best agreement with the ground truth.

Then, we continue with the Bayesian test, \cite{bentivegnaBayesianApproachBoson2015}, where two hypothetical theoretical models are compared against each other based on the likelihood to generate the experimental samples. We define the Bayesian test score $\Delta H$ as the difference between the ground-truth model $H_0$ and the classical adversary model hypothesis $H_1$ on a set of $n$-photon-click samples:
\begin{equation}
    \Delta{H}={\frac{1}{N}}\mathrm{ln}\prod_{i=1}^{N}{\frac{{{P}}^{(0)}({\vec{{s}}}_{i}){{P}}^{(1)}(n)}{{P}^{(1)}({\vec{{s}}}_{i}){{P}}^{(0)}(n)}}
\end{equation}
where $N$ is the number of samples, $\vec{s}_i$ represents the $i$th sample, ${P}^{(0/1)}({\vec{{s}}}_{i})$ represents the event probability of sample $\vec{s}_i$ under the $H_{0/1}$ hypothesis , and ${P}^{(0/1)}(n)$ is the coarse-grained probability for the $n$-photon-click under hypothesis $H_{0/1}$. When $\Delta H > 0$, it is proven that the experimental samples are more likely from the ground-truth GBS rather than the mockup. A higher Bayesian test score indicates larger confidence. 

For the validation, each of the 144 modes is treated as 1-to-8 fan out, yielding a total mode number of 1152 for the ground truth. Further, in our model we consider the noise of the partial photon distinguishability for the Bayesian test, where the probability of the sample event is calculated by a modified version of the Torontonian \cite{suppl}. We find that our model gives a closer description of the experiment than the previous ones. We use the \textit{Sunway TaihuLight} supercomputer to calculate the probability of large photon number samples in the QCA regime.

The Bayesian test score of the ground truth against the mockups using the thermal state and the squashed state at \SI{1.3}{W} pump power are shown in Fig. \ref{fig:2}(c)-\ref{fig:2}(d). Due to the classical computational overhead, we first start from a subsystem with fewer output bosonic modes, and gradually increase the subsystem size. We observe not only all the Bayesian scores in Fig. \ref{fig:2}(c)-\ref{fig:2}(d) are higher than zero, but also they show an evident increasing trend as the subsystem size ramps up.
The positive scores demonstrate that the experimental GBS samples are more likely generated from the ground-truth distribution rather than these mockup distributions. More importantly, the rising trend of the Bayesian score indicates a higher score can be inferred for the full system. This increasing trend could be understood from that, as subsystem size increases, more complete information of the full system is incorporated into the Bayesian test, and therefore results in stronger validation strength.
Therefore, we conclusively infer that for the full-mode system, though it is not directly computable, stronger Bayesian confidence for the ground-truth theory is expected.

We continue to investigate the power dependence of the Bayesian test. In Fig. \ref{fig:2}(e)-\ref{fig:2}(f), the Bayesian validation strength is plotted for three different pump laser powers from \SI{0.72}{W} to \SI{1.3}{W}. The results show that the lower pump power can generate higher Bayesian score. This is expected because the thermal noise and photon loss become increasingly sensitive with larger squeezing parameters \cite{qiRegimesClassicalSimulability2020}.

Another important tool for characterizing the GBS is the correlation function. The $k$-order correlation function is recursively defined as
\begin{equation}
    \kappa(X_{1}X_{2}\cdots X_{k})=E(X_{1}X_{2}\cdots X_{k})-\sum_{p\in P_k}\prod_{b\in p}\kappa[(X_{i})_{i\in b}]
\end{equation}
where $X_k$ represents the experimental measurement operator at the $k$th output mode, and $P_k$ represent all partitions of the set $\{1,2,\cdots,k\}$ excluding the universal set. Based on the correlation function, we validate our experimental samples against other recently proposed mockups based on approximating the experimental GBS through low-order marginal distributions \cite{villalongaEfficientApproximationExperimental2022}.

Reference \cite{villalongaEfficientApproximationExperimental2022} proposed a greedy algorithm to sample from a distribution that approximates all first- and second-order correlations of the experiment, which we call a greedy sampler below. The same work also used the single-mode marginals and two-mode correlations to configure a Boltzmann machine for sampling under the Thouless, Anderson, and Palmer (TAP) mean field approximation. These methods can be generalized to higher orders in principle, but a full enumeration of all the marginal probabilities of the chosen order is required, which grow combinatorically, and thus limits it within the low-order regime. However, these mockups do not capture higher-order correlations, and can thus be ruled out from this aspect.

In Fig. \ref{fig:3}(a)-\ref{fig:3}(b), we directly compare the third-order correlation functions between the experimental samples at the QCA regime and the mockup samplers from the order-2 greedy and the order-2 TAP samplers. The statistics of the experimental samples agree with the ground truth where they cluster around the identity line at \SI{45}{\degree}, whereas the mockup samplers’ third-order correlations show significant divergence from the ground-truth predictions.

Additionally, Ref. \cite{bulmerBoundaryQuantumAdvantage2022} designed the IPS sampler, which was shown to yield higher heavy output generation (HOG) score than the mockups using the thermal states and distinguishable photons. We compare the second-order cumulants of the sampler with the experimental results in Fig. \ref{fig:3}(c), where the IPS sampler shows an evident deviation from the results of the ground truth and the experimental data.

\begin{figure}[tbp]
    \includegraphics[width = 0.45\textwidth]{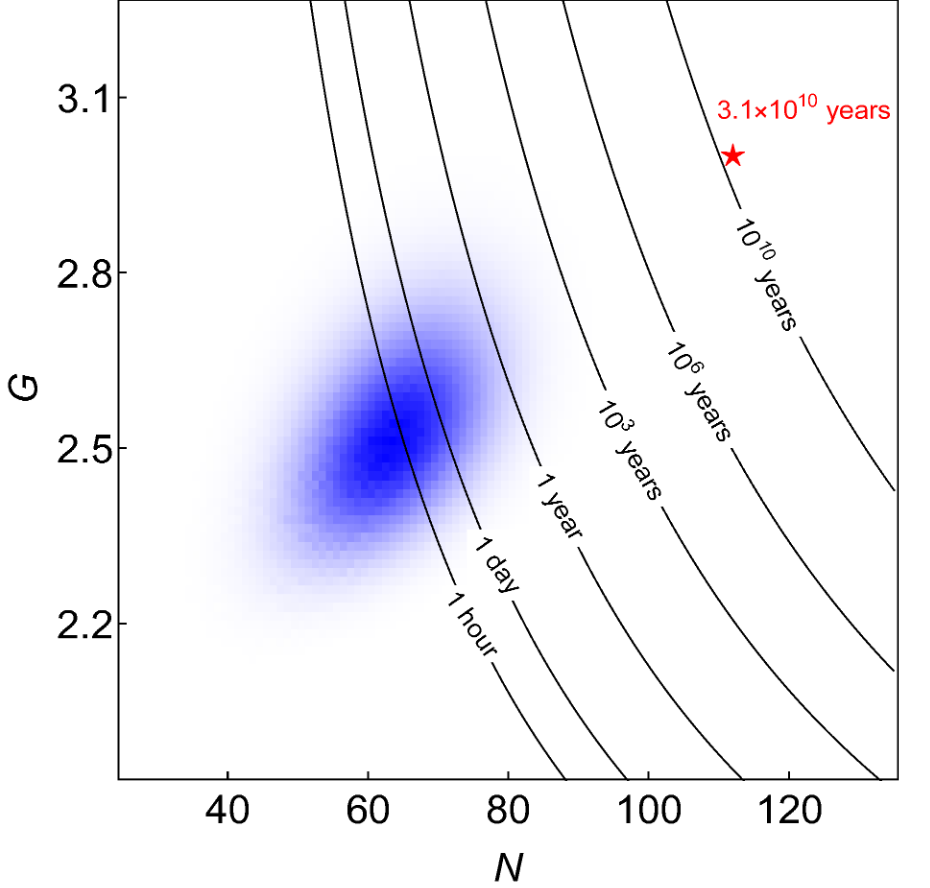}
    \caption{Quantum computational advantage. The distribution of the estimated classical simulation time on \textit{Frontier} with exact methods. We calculate the paired value $(G,N)$ for each of the samples obtained in the highest power experiment, and construct the two-dimensional heat map. $G>2$ indicates the feature of photon number resolving. The hardest sample is marked in star. Equal simulation time given by equation (\ref{eqn:3}) is shown as the lines. As a comparison, it takes \SI{1.27}{\us} to produce a sample in our experiment.
    }
    \label{fig:4}
\end{figure}

The last mockup to be ruled out is the recently proposed treewidth sampler \cite{ohClassicalSimulationBoson2022} which can use local connectivity of the circuits to reduce the overhead of classical approximate sampling. The sampler could generate higher HOG score than some GBS experiments, which, however, is due to the limitation of HOG test itself. In fact, one can show that the treewidth mockup sampler can even yield a higher HOG score than an ideal ground-truth sampler (see Supplementary Materials \cite{suppl}), therefore makes the test unreliable \cite{ohSpoofingCrossEntropy2022} (related arguments on the limitation of the cross-entropy benchmarking in random circuit sampling have also been reported \cite{barak2020spoofing, gaoLimitationsLinearCrossEntropy2021}). Thus, it is necessary to rule out the treewidth mockup in other ways. In Fig. \ref{fig:3}(d), we show the second-order correlation of a treewidth mockup (with a propagation length of 65) and the experimental sampler. The former deviates significantly from the ground-truth theory, and is thus unambiguously ruled out.

As a complementary test, we also show in \ref{fig:3}(e) the second order correlation of the experiment, the thermal state and the squashed state mockup. The correlation of the experiment, with two-norm distance $D=0.040$ and slope $K=1.006$, clearly agrees with the ground truth better the thermal state mockup ($D=0.275, K=2.108$) and the squashed mockup ($D=0.052, K=1.110$).

Having excluded all currently proposed mockups, we now benchmark the classical computational cost to simulate a noiseless version of our GBS experiment. Our pseudo-photon number resolving detection scheme can be modelled by treating all the 1152 fan-out bins as individual output modes, each with threshold detection. We can use results from \cite{bulmerBoundaryQuantumAdvantage2022} to exploit the fact that different output bins corresponding to the same optical mode result in repeated entries in the loop hafnian during the simulation, further reducing the dominant computational cost to:
\begin{equation}
    T(\vec{N})=\frac{1}{2}C_{\rm{F r o n t i e r}}M N^{3}G^{N/2}
    \label{eqn:3}
\end{equation}
where $\vec{N}=\{n_1,n_2,\cdots,n_{144}\}$ represents the PPNRD sample with $n_i$ being the photon-click number of the  $i$th mode, $N$ is the number of clicked modes, $M$ is the mode number, $G=\left(\Pi_{i}^{M}(n_{i}+1)\right)^{1/N}$, and $C_{\rm{F r o n t i e r}}$ is estimated based on \cite{madsenQuantumComputationalAdvantage2022}. The addition of PPNRD therefore substantially increases the computational complexity of our experiment compared to threshold detection (which always has $G=2$), due to the strong increase in the number of registered detection events.

We now estimate the time cost on \textit{Frontier}, currently the most powerful supercomputer. For each sample of our high power experiment, we estimate it would on average take \textit{Frontier} at least $\sim 600$ years to generate using the exact methods, while it only takes our machine \SI{1.27}{\us} to produce a sample, showing an overwhelming QCA of \num{1.5e16}. Moreover, the hardest sample from our experiment would take \textit{Frontier} more than $\sim$ \num{3.1e10} years to generate using exact algorithm. We show in Fig. \ref{fig:4} the distribution of the estimated classical overhead of all the experimental samples. We hope future work will further consider the realistic experimental imperfection such as photon loss and partial photon distinguishability for a better benchmark \cite{oh2023newsim}.

\begin{acknowledgments}

We thank Changhun Oh, Minzhao Liu and Liang Jiang for interesting discussions. We thank Z.-M. He, R.-X. Wang, X.-Y. Wu, Z.-M. Zhang and S.-T. Zheng for helpful assistance during the experiment. We thank Y. Liang and Q.-H. Shi for their help in preparation of the manuscript. This work was supported by the National Key R\&D Program of China (2019YFA0308700), the National Natural Science Foundation of China, the Chinese Academy of Sciences, the Anhui Initiative in Quantum Information Technologies, the Science and Technology Commission of Shanghai Municipality(2019SHZDZX01), the Innovation Program for Quantum Science and Technology (No. ZD0202010000), the XPLORER PRIZE, the New Cornerstone Science Foundation, and NWO Veni.

Y.-H. D., Y.-C. G., H.-L. L., S.-Q. G. and H. S. contributed equally to
this work.

The dataset of this experiment can be accessed via \cite{JZ3data}.

\textit{Note added} - After submission, we became aware of a recent related theoretical proposal for balanced-fan-out PPNRD GBS \cite{bressanini2023PPNRDgaussian}.

\end{acknowledgments}


\begin{thebibliography}{54}%
\makeatletter
\providecommand \@ifxundefined [1]{%
 \@ifx{#1\undefined}
}%
\providecommand \@ifnum [1]{%
 \ifnum #1\expandafter \@firstoftwo
 \else \expandafter \@secondoftwo
 \fi
}%
\providecommand \@ifx [1]{%
 \ifx #1\expandafter \@firstoftwo
 \else \expandafter \@secondoftwo
 \fi
}%
\providecommand \natexlab [1]{#1}%
\providecommand \enquote  [1]{``#1''}%
\providecommand \bibnamefont  [1]{#1}%
\providecommand \bibfnamefont [1]{#1}%
\providecommand \citenamefont [1]{#1}%
\providecommand \href@noop [0]{\@secondoftwo}%
\providecommand \href [0]{\begingroup \@sanitize@url \@href}%
\providecommand \@href[1]{\@@startlink{#1}\@@href}%
\providecommand \@@href[1]{\endgroup#1\@@endlink}%
\providecommand \@sanitize@url [0]{\catcode `\\12\catcode `\$12\catcode
  `\&12\catcode `\#12\catcode `\^12\catcode `\_12\catcode `\%12\relax}%
\providecommand \@@startlink[1]{}%
\providecommand \@@endlink[0]{}%
\providecommand \url  [0]{\begingroup\@sanitize@url \@url }%
\providecommand \@url [1]{\endgroup\@href {#1}{\urlprefix }}%
\providecommand \urlprefix  [0]{URL }%
\providecommand \Eprint [0]{\href }%
\providecommand \doibase [0]{https://doi.org/}%
\providecommand \selectlanguage [0]{\@gobble}%
\providecommand \bibinfo  [0]{\@secondoftwo}%
\providecommand \bibfield  [0]{\@secondoftwo}%
\providecommand \translation [1]{[#1]}%
\providecommand \BibitemOpen [0]{}%
\providecommand \bibitemStop [0]{}%
\providecommand \bibitemNoStop [0]{.\EOS\space}%
\providecommand \EOS [0]{\spacefactor3000\relax}%
\providecommand \BibitemShut  [1]{\csname bibitem#1\endcsname}%
\let\auto@bib@innerbib\@empty
\bibitem [{\citenamefont {Bernstein}\ and\ \citenamefont
  {Vazirani}(1993)}]{bernsteinQuantumComplexityTheory1993}%
  \BibitemOpen
  \bibfield  {author} {\bibinfo {author} {\bibfnamefont {E.}~\bibnamefont
  {Bernstein}}\ and\ \bibinfo {author} {\bibfnamefont {U.}~\bibnamefont
  {Vazirani}},\ }in\ \href {https://doi.org/10.1145/167088.167097} {\emph
  {\bibinfo {booktitle} {Proceedings of the Twenty-Fifth Annual {{ACM}}
  Symposium on {{Theory}} of Computing - {{STOC}} '93}}}\ (\bibinfo
  {publisher} {{ACM Press}},\ \bibinfo {address} {{San Diego, California,
  United States}},\ \bibinfo {year} {1993})\ pp.\ \bibinfo {pages}
  {11--20}\BibitemShut {NoStop}%
\bibitem [{\citenamefont
  {Preskill}(2012)}]{preskillQuantumComputingEntanglement2012}%
  \BibitemOpen
  \bibfield  {author} {\bibinfo {author} {\bibfnamefont {J.}~\bibnamefont
  {Preskill}},\ }\href {http://arxiv.org/abs/1203.5813} {\bibfield  {journal}
  {\bibinfo  {journal} {arXiv:1203.5813}\ } (\bibinfo {year}
  {2012})}\BibitemShut {NoStop}%
\bibitem [{\citenamefont {Harrow}\ and\ \citenamefont
  {Montanaro}(2017)}]{harrowQuantumComputationalSupremacy2017}%
  \BibitemOpen
  \bibfield  {author} {\bibinfo {author} {\bibfnamefont {A.~W.}\ \bibnamefont
  {Harrow}}\ and\ \bibinfo {author} {\bibfnamefont {A.}~\bibnamefont
  {Montanaro}},\ }\href {https://doi.org/10.1038/nature23458} {\bibfield
  {journal} {\bibinfo  {journal} {Nature}\ }\textbf {\bibinfo {volume} {549}},\
  \bibinfo {pages} {203} (\bibinfo {year} {2017})}\BibitemShut {NoStop}%
\bibitem [{\citenamefont {Lund}\ \emph {et~al.}(2017)\citenamefont {Lund},
  \citenamefont {Bremner},\ and\ \citenamefont
  {Ralph}}]{lundQuantumSamplingProblems2017}%
  \BibitemOpen
  \bibfield  {author} {\bibinfo {author} {\bibfnamefont {A.~P.}\ \bibnamefont
  {Lund}}, \bibinfo {author} {\bibfnamefont {M.~J.}\ \bibnamefont {Bremner}},\
  and\ \bibinfo {author} {\bibfnamefont {T.~C.}\ \bibnamefont {Ralph}},\ }\href
  {https://doi.org/10.1038/s41534-017-0018-2} {\bibfield  {journal} {\bibinfo
  {journal} {npj Quantum Information}\ }\textbf {\bibinfo {volume} {3}},\
  \bibinfo {pages} {15} (\bibinfo {year} {2017})}\BibitemShut {NoStop}%
\bibitem [{\citenamefont {Hangleiter}\ and\ \citenamefont
  {Eisert}(2023)}]{hangleiterComputationalAdvantageQuantum2022}%
  \BibitemOpen
  \bibfield  {author} {\bibinfo {author} {\bibfnamefont {D.}~\bibnamefont
  {Hangleiter}}\ and\ \bibinfo {author} {\bibfnamefont {J.}~\bibnamefont
  {Eisert}},\ }\href@noop {} {\bibfield  {journal} {\bibinfo  {journal}
  {arXiv:2206.04079}\ } (\bibinfo {year} {2023})}\BibitemShut {NoStop}%
\bibitem [{\citenamefont {Aaronson}\ and\ \citenamefont
  {Arkhipov}(2011)}]{Aaronson2011}%
  \BibitemOpen
  \bibfield  {author} {\bibinfo {author} {\bibfnamefont {S.}~\bibnamefont
  {Aaronson}}\ and\ \bibinfo {author} {\bibfnamefont {A.}~\bibnamefont
  {Arkhipov}},\ }in\ \href {https://doi.org/10.1145/1993636.1993682} {\emph
  {\bibinfo {booktitle} {Proceedings of the 43rd Annual {{ACM}} Symposium on
  {{Theory}} of Computing - {{STOC}} '11}}}\ (\bibinfo  {publisher} {{ACM
  Press}},\ \bibinfo {address} {{New York, New York, USA}},\ \bibinfo {year}
  {2011})\ pp.\ \bibinfo {pages} {333--342}\BibitemShut {NoStop}%
\bibitem [{\citenamefont {Boixo}\ \emph {et~al.}(2018)\citenamefont {Boixo},
  \citenamefont {Isakov}, \citenamefont {Smelyanskiy}, \citenamefont {Babbush},
  \citenamefont {Ding}, \citenamefont {Jiang}, \citenamefont {Bremner},
  \citenamefont {Martinis},\ and\ \citenamefont
  {Neven}}]{boixo2018characterizing}%
  \BibitemOpen
  \bibfield  {author} {\bibinfo {author} {\bibfnamefont {S.}~\bibnamefont
  {Boixo}}, \bibinfo {author} {\bibfnamefont {S.~V.}\ \bibnamefont {Isakov}},
  \bibinfo {author} {\bibfnamefont {V.~N.}\ \bibnamefont {Smelyanskiy}},
  \bibinfo {author} {\bibfnamefont {R.}~\bibnamefont {Babbush}}, \bibinfo
  {author} {\bibfnamefont {N.}~\bibnamefont {Ding}}, \bibinfo {author}
  {\bibfnamefont {Z.}~\bibnamefont {Jiang}}, \bibinfo {author} {\bibfnamefont
  {M.~J.}\ \bibnamefont {Bremner}}, \bibinfo {author} {\bibfnamefont {J.~M.}\
  \bibnamefont {Martinis}},\ and\ \bibinfo {author} {\bibfnamefont
  {H.}~\bibnamefont {Neven}},\ }\href
  {https://doi.org/10.1038/s41567-018-0124-x} {\bibfield  {journal} {\bibinfo
  {journal} {Nature Physics}\ }\textbf {\bibinfo {volume} {14}},\ \bibinfo
  {pages} {595} (\bibinfo {year} {2018})}\BibitemShut {NoStop}%
\bibitem [{\citenamefont {Arute}\ \emph {et~al.}(2019)\citenamefont {Arute},
  \citenamefont {Arya}, \citenamefont {Babbush}, \citenamefont {Bacon},
  \citenamefont {Bardin} \emph {et~al.}}]{aruteQuantumSupremacyUsing2019}%
  \BibitemOpen
  \bibfield  {author} {\bibinfo {author} {\bibfnamefont {F.}~\bibnamefont
  {Arute}}, \bibinfo {author} {\bibfnamefont {K.}~\bibnamefont {Arya}},
  \bibinfo {author} {\bibfnamefont {R.}~\bibnamefont {Babbush}}, \bibinfo
  {author} {\bibfnamefont {D.}~\bibnamefont {Bacon}}, \bibinfo {author}
  {\bibfnamefont {J.~C.}\ \bibnamefont {Bardin}}, \emph {et~al.},\ }\href
  {https://doi.org/10.1038/s41586-019-1666-5} {\bibfield  {journal} {\bibinfo
  {journal} {Nature}\ }\textbf {\bibinfo {volume} {574}},\ \bibinfo {pages}
  {505} (\bibinfo {year} {2019})}\BibitemShut {NoStop}%
\bibitem [{\citenamefont {Zhong}\ \emph
  {et~al.}(2021{\natexlab{a}})\citenamefont {Zhong}, \citenamefont {Wang},
  \citenamefont {Deng}, \citenamefont {Chen}, \citenamefont {Peng} \emph
  {et~al.}}]{Zhong2021}%
  \BibitemOpen
  \bibfield  {author} {\bibinfo {author} {\bibfnamefont {H.~S.}\ \bibnamefont
  {Zhong}}, \bibinfo {author} {\bibfnamefont {H.}~\bibnamefont {Wang}},
  \bibinfo {author} {\bibfnamefont {Y.~H.}\ \bibnamefont {Deng}}, \bibinfo
  {author} {\bibfnamefont {M.~C.}\ \bibnamefont {Chen}}, \bibinfo {author}
  {\bibfnamefont {L.~C.}\ \bibnamefont {Peng}}, \emph {et~al.},\ }\href
  {https://doi.org/10.1126/science.abe8770} {\bibfield  {journal} {\bibinfo
  {journal} {Science}\ }\textbf {\bibinfo {volume} {370}},\ \bibinfo {pages}
  {1460} (\bibinfo {year} {2021}{\natexlab{a}})}\BibitemShut {NoStop}%
\bibitem [{\citenamefont {Zhong}\ \emph
  {et~al.}(2021{\natexlab{b}})\citenamefont {Zhong}, \citenamefont {Deng},
  \citenamefont {Qin}, \citenamefont {Wang}, \citenamefont {Chen} \emph
  {et~al.}}]{zhongPhaseProgrammableGaussianBoson2021}%
  \BibitemOpen
  \bibfield  {author} {\bibinfo {author} {\bibfnamefont {H.-S.}\ \bibnamefont
  {Zhong}}, \bibinfo {author} {\bibfnamefont {Y.-H.}\ \bibnamefont {Deng}},
  \bibinfo {author} {\bibfnamefont {J.}~\bibnamefont {Qin}}, \bibinfo {author}
  {\bibfnamefont {H.}~\bibnamefont {Wang}}, \bibinfo {author} {\bibfnamefont
  {M.-C.}\ \bibnamefont {Chen}}, \emph {et~al.},\ }\href
  {https://doi.org/10.1103/PhysRevLett.127.180502} {\bibfield  {journal}
  {\bibinfo  {journal} {Physical Review Letters}\ }\textbf {\bibinfo {volume}
  {127}},\ \bibinfo {pages} {180502} (\bibinfo {year}
  {2021}{\natexlab{b}})}\BibitemShut {NoStop}%
\bibitem [{\citenamefont {Wu}\ \emph {et~al.}(2021)\citenamefont {Wu},
  \citenamefont {Bao}, \citenamefont {Cao}, \citenamefont {Chen}, \citenamefont
  {Chen} \emph {et~al.}}]{wuStrongQuantumComputational2021}%
  \BibitemOpen
  \bibfield  {author} {\bibinfo {author} {\bibfnamefont {Y.}~\bibnamefont
  {Wu}}, \bibinfo {author} {\bibfnamefont {W.-S.}\ \bibnamefont {Bao}},
  \bibinfo {author} {\bibfnamefont {S.}~\bibnamefont {Cao}}, \bibinfo {author}
  {\bibfnamefont {F.}~\bibnamefont {Chen}}, \bibinfo {author} {\bibfnamefont
  {M.-C.}\ \bibnamefont {Chen}}, \emph {et~al.},\ }\href
  {https://doi.org/10.1103/PhysRevLett.127.180501} {\bibfield  {journal}
  {\bibinfo  {journal} {Physical Review Letters}\ }\textbf {\bibinfo {volume}
  {127}},\ \bibinfo {pages} {180501} (\bibinfo {year} {2021})}\BibitemShut
  {NoStop}%
\bibitem [{\citenamefont {Madsen}\ \emph {et~al.}(2022)\citenamefont {Madsen},
  \citenamefont {Laudenbach}, \citenamefont {Askarani}, \citenamefont
  {Rortais}, \citenamefont {Vincent}, \citenamefont {Bulmer} \emph
  {et~al.}}]{madsenQuantumComputationalAdvantage2022}%
  \BibitemOpen
  \bibfield  {author} {\bibinfo {author} {\bibfnamefont {L.~S.}\ \bibnamefont
  {Madsen}}, \bibinfo {author} {\bibfnamefont {F.}~\bibnamefont {Laudenbach}},
  \bibinfo {author} {\bibfnamefont {M.~F.}\ \bibnamefont {Askarani}}, \bibinfo
  {author} {\bibfnamefont {F.}~\bibnamefont {Rortais}}, \bibinfo {author}
  {\bibfnamefont {T.}~\bibnamefont {Vincent}}, \bibinfo {author} {\bibfnamefont
  {J.~F.~F.}\ \bibnamefont {Bulmer}}, \emph {et~al.},\ }\href
  {https://doi.org/10.1038/s41586-022-04725-x} {\bibfield  {journal} {\bibinfo
  {journal} {Nature}\ }\textbf {\bibinfo {volume} {606}},\ \bibinfo {pages}
  {75} (\bibinfo {year} {2022})}\BibitemShut {NoStop}%
\bibitem [{\citenamefont {Bell}(1964)}]{bellEinsteinPodolskyRosen1964}%
  \BibitemOpen
  \bibfield  {author} {\bibinfo {author} {\bibfnamefont {J.~S.}\ \bibnamefont
  {Bell}},\ }\href {https://doi.org/10.1103/PhysicsPhysiqueFizika.1.195}
  {\bibfield  {journal} {\bibinfo  {journal} {Physics Physique Fizika}\
  }\textbf {\bibinfo {volume} {1}},\ \bibinfo {pages} {195} (\bibinfo {year}
  {1964})}\BibitemShut {NoStop}%
\bibitem [{\citenamefont {Einstein}\ \emph {et~al.}(1935)\citenamefont
  {Einstein}, \citenamefont {Podolsky},\ and\ \citenamefont
  {Rosen}}]{einsteinCanQuantumMechanicalDescription1935}%
  \BibitemOpen
  \bibfield  {author} {\bibinfo {author} {\bibfnamefont {A.}~\bibnamefont
  {Einstein}}, \bibinfo {author} {\bibfnamefont {B.}~\bibnamefont {Podolsky}},\
  and\ \bibinfo {author} {\bibfnamefont {N.}~\bibnamefont {Rosen}},\ }\href
  {https://doi.org/10.1103/PhysRev.47.777} {\bibfield  {journal} {\bibinfo
  {journal} {Physical Review}\ }\textbf {\bibinfo {volume} {47}},\ \bibinfo
  {pages} {777} (\bibinfo {year} {1935})}\BibitemShut {NoStop}%
\bibitem [{\citenamefont {Ágoston Kaposi}\ \emph {et~al.}(2022)\citenamefont
  {Ágoston Kaposi}, \citenamefont {Kolarovszki}, \citenamefont {Kozsik},
  \citenamefont {Zimborás},\ and\ \citenamefont
  {Rakyta}}]{kaposiPolynomialSpeedupTorontonian2021}%
  \BibitemOpen
  \bibfield  {author} {\bibinfo {author} {\bibnamefont {Ágoston Kaposi}},
  \bibinfo {author} {\bibfnamefont {Z.}~\bibnamefont {Kolarovszki}}, \bibinfo
  {author} {\bibfnamefont {T.}~\bibnamefont {Kozsik}}, \bibinfo {author}
  {\bibfnamefont {Z.}~\bibnamefont {Zimborás}},\ and\ \bibinfo {author}
  {\bibfnamefont {P.}~\bibnamefont {Rakyta}},\ }\href@noop {} {\bibfield
  {journal} {\bibinfo  {journal} {arXiv:2109.04528}\ } (\bibinfo {year}
  {2022})}\BibitemShut {NoStop}%
\bibitem [{\citenamefont {Quesada}\ \emph {et~al.}(2022)\citenamefont
  {Quesada}, \citenamefont {Chadwick}, \citenamefont {Bell}, \citenamefont
  {Arrazola}, \citenamefont {Vincent}, \citenamefont {Qi},\ and\ \citenamefont
  {Garc{\'i}a-Patr{\'o}n}}]{quesadaQuadraticSpeedUpSimulating2022}%
  \BibitemOpen
  \bibfield  {author} {\bibinfo {author} {\bibfnamefont {N.}~\bibnamefont
  {Quesada}}, \bibinfo {author} {\bibfnamefont {R.~S.}\ \bibnamefont
  {Chadwick}}, \bibinfo {author} {\bibfnamefont {B.~A.}\ \bibnamefont {Bell}},
  \bibinfo {author} {\bibfnamefont {J.~M.}\ \bibnamefont {Arrazola}}, \bibinfo
  {author} {\bibfnamefont {T.}~\bibnamefont {Vincent}}, \bibinfo {author}
  {\bibfnamefont {H.}~\bibnamefont {Qi}},\ and\ \bibinfo {author}
  {\bibfnamefont {R.}~\bibnamefont {Garc{\'i}a-Patr{\'o}n}},\ }\href
  {https://doi.org/10.1103/PRXQuantum.3.010306} {\bibfield  {journal} {\bibinfo
   {journal} {PRX Quantum}\ }\textbf {\bibinfo {volume} {3}},\ \bibinfo {pages}
  {010306} (\bibinfo {year} {2022})}\BibitemShut {NoStop}%
\bibitem [{\citenamefont {Bulmer}\ \emph {et~al.}(2022)\citenamefont {Bulmer},
  \citenamefont {Bell}, \citenamefont {Chadwick}, \citenamefont {Jones},
  \citenamefont {Moise} \emph {et~al.}}]{bulmerBoundaryQuantumAdvantage2022}%
  \BibitemOpen
  \bibfield  {author} {\bibinfo {author} {\bibfnamefont {J.~F.~F.}\
  \bibnamefont {Bulmer}}, \bibinfo {author} {\bibfnamefont {B.~A.}\
  \bibnamefont {Bell}}, \bibinfo {author} {\bibfnamefont {R.~S.}\ \bibnamefont
  {Chadwick}}, \bibinfo {author} {\bibfnamefont {A.~E.}\ \bibnamefont {Jones}},
  \bibinfo {author} {\bibfnamefont {D.}~\bibnamefont {Moise}}, \emph {et~al.},\
  }\href {https://doi.org/10.1126/sciadv.abl9236} {\bibfield  {journal}
  {\bibinfo  {journal} {Science Advances}\ }\textbf {\bibinfo {volume} {8}},\
  \bibinfo {pages} {eabl9236} (\bibinfo {year} {2022})}\BibitemShut {NoStop}%
\bibitem [{\citenamefont {Qi}\ \emph {et~al.}(2020)\citenamefont {Qi},
  \citenamefont {Brod}, \citenamefont {Quesada},\ and\ \citenamefont
  {{Garc{\'i}a-Patr{\'o}n}}}]{qiRegimesClassicalSimulability2020}%
  \BibitemOpen
  \bibfield  {author} {\bibinfo {author} {\bibfnamefont {H.}~\bibnamefont
  {Qi}}, \bibinfo {author} {\bibfnamefont {D.~J.}\ \bibnamefont {Brod}},
  \bibinfo {author} {\bibfnamefont {N.}~\bibnamefont {Quesada}},\ and\ \bibinfo
  {author} {\bibfnamefont {R.}~\bibnamefont {{Garc{\'i}a-Patr{\'o}n}}},\ }\href
  {https://doi.org/10.1103/PhysRevLett.124.100502} {\bibfield  {journal}
  {\bibinfo  {journal} {Physical Review Letters}\ }\textbf {\bibinfo {volume}
  {124}},\ \bibinfo {pages} {100502} (\bibinfo {year} {2020})}\BibitemShut
  {NoStop}%
\bibitem [{\citenamefont
  {Renema}(2020)}]{renemaSimulabilityPartiallyDistinguishable2020}%
  \BibitemOpen
  \bibfield  {author} {\bibinfo {author} {\bibfnamefont {J.~J.}\ \bibnamefont
  {Renema}},\ }\href {https://doi.org/10.1103/PhysRevA.101.063840} {\bibfield
  {journal} {\bibinfo  {journal} {Physical Review A}\ }\textbf {\bibinfo
  {volume} {101}},\ \bibinfo {pages} {063840} (\bibinfo {year}
  {2020})}\BibitemShut {NoStop}%
\bibitem [{\citenamefont {Shi}\ and\ \citenamefont
  {Byrnes}(2022)}]{shiEffectPartialDistinguishability2022}%
  \BibitemOpen
  \bibfield  {author} {\bibinfo {author} {\bibfnamefont {J.}~\bibnamefont
  {Shi}}\ and\ \bibinfo {author} {\bibfnamefont {T.}~\bibnamefont {Byrnes}},\
  }\href {https://doi.org/10.1038/s41534-022-00557-9} {\bibfield  {journal}
  {\bibinfo  {journal} {npj Quantum Information}\ }\textbf {\bibinfo {volume}
  {8}},\ \bibinfo {pages} {54} (\bibinfo {year} {2022})}\BibitemShut {NoStop}%
\bibitem [{\citenamefont {Villalonga}\ \emph {et~al.}(2022)\citenamefont
  {Villalonga}, \citenamefont {Niu}, \citenamefont {Li}, \citenamefont {Neven},
  \citenamefont {Platt}, \citenamefont {Smelyanskiy},\ and\ \citenamefont
  {Boixo}}]{villalongaEfficientApproximationExperimental2022}%
  \BibitemOpen
  \bibfield  {author} {\bibinfo {author} {\bibfnamefont {B.}~\bibnamefont
  {Villalonga}}, \bibinfo {author} {\bibfnamefont {M.~Y.}\ \bibnamefont {Niu}},
  \bibinfo {author} {\bibfnamefont {L.}~\bibnamefont {Li}}, \bibinfo {author}
  {\bibfnamefont {H.}~\bibnamefont {Neven}}, \bibinfo {author} {\bibfnamefont
  {J.~C.}\ \bibnamefont {Platt}}, \bibinfo {author} {\bibfnamefont {V.~N.}\
  \bibnamefont {Smelyanskiy}},\ and\ \bibinfo {author} {\bibfnamefont
  {S.}~\bibnamefont {Boixo}},\ }\href {http://arxiv.org/abs/2109.11525}
  {\bibfield  {journal} {\bibinfo  {journal} {arXiv:2109.11525}\ } (\bibinfo
  {year} {2022})}\BibitemShut {NoStop}%
\bibitem [{\citenamefont {Oh}\ \emph {et~al.}(2022{\natexlab{a}})\citenamefont
  {Oh}, \citenamefont {Lim}, \citenamefont {Fefferman},\ and\ \citenamefont
  {Jiang}}]{ohClassicalSimulationBoson2022}%
  \BibitemOpen
  \bibfield  {author} {\bibinfo {author} {\bibfnamefont {C.}~\bibnamefont
  {Oh}}, \bibinfo {author} {\bibfnamefont {Y.}~\bibnamefont {Lim}}, \bibinfo
  {author} {\bibfnamefont {B.}~\bibnamefont {Fefferman}},\ and\ \bibinfo
  {author} {\bibfnamefont {L.}~\bibnamefont {Jiang}},\ }\href
  {http://arxiv.org/abs/2110.01564} {\bibfield  {journal} {\bibinfo  {journal}
  {arXiv:2110.01564}\ } (\bibinfo {year} {2022}{\natexlab{a}})}\BibitemShut
  {NoStop}%
\bibitem [{\citenamefont {Oh}\ \emph {et~al.}(2022{\natexlab{b}})\citenamefont
  {Oh}, \citenamefont {Jiang},\ and\ \citenamefont
  {Fefferman}}]{ohSpoofingCrossEntropy2022}%
  \BibitemOpen
  \bibfield  {author} {\bibinfo {author} {\bibfnamefont {C.}~\bibnamefont
  {Oh}}, \bibinfo {author} {\bibfnamefont {L.}~\bibnamefont {Jiang}},\ and\
  \bibinfo {author} {\bibfnamefont {B.}~\bibnamefont {Fefferman}},\ }\href
  {http://arxiv.org/abs/2210.15021} {\bibfield  {journal} {\bibinfo  {journal}
  {arXiv:2210.15021}\ } (\bibinfo {year} {2022}{\natexlab{b}})}\BibitemShut
  {NoStop}%
\bibitem [{\citenamefont
  {Shchesnovich}(2021)}]{shchesnovich2021distinguishing}%
  \BibitemOpen
  \bibfield  {author} {\bibinfo {author} {\bibfnamefont {V.}~\bibnamefont
  {Shchesnovich}},\ }\href {https://doi.org/10.22331/q-2021-03-29-423}
  {\bibfield  {journal} {\bibinfo  {journal} {Quantum}\ }\textbf {\bibinfo
  {volume} {5}},\ \bibinfo {pages} {423} (\bibinfo {year} {2021})}\BibitemShut
  {NoStop}%
\bibitem [{\citenamefont {Dellios}\ \emph
  {et~al.}(2022{\natexlab{a}})\citenamefont {Dellios}, \citenamefont {Reid},
  \citenamefont {Opanchuk},\ and\ \citenamefont
  {Drummond}}]{delliosValidationTestsGBS2022}%
  \BibitemOpen
  \bibfield  {author} {\bibinfo {author} {\bibfnamefont {A.~S.}\ \bibnamefont
  {Dellios}}, \bibinfo {author} {\bibfnamefont {M.~D.}\ \bibnamefont {Reid}},
  \bibinfo {author} {\bibfnamefont {B.}~\bibnamefont {Opanchuk}},\ and\
  \bibinfo {author} {\bibfnamefont {P.~D.}\ \bibnamefont {Drummond}},\ }\href
  {http://arxiv.org/abs/2211.03480} {\bibfield  {journal} {\bibinfo  {journal}
  {arXiv:2211.03480}\ } (\bibinfo {year} {2022}{\natexlab{a}})}\BibitemShut
  {NoStop}%
\bibitem [{\citenamefont {Seron}\ \emph {et~al.}(2022)\citenamefont {Seron},
  \citenamefont {Novo}, \citenamefont {Arkhipov},\ and\ \citenamefont
  {Cerf}}]{seron2022efficient}%
  \BibitemOpen
  \bibfield  {author} {\bibinfo {author} {\bibfnamefont {B.}~\bibnamefont
  {Seron}}, \bibinfo {author} {\bibfnamefont {L.}~\bibnamefont {Novo}},
  \bibinfo {author} {\bibfnamefont {A.}~\bibnamefont {Arkhipov}},\ and\
  \bibinfo {author} {\bibfnamefont {N.~J.}\ \bibnamefont {Cerf}},\ }\href
  {https://doi.org/10.48550/arXiv.2212.09643} {\bibfield  {journal} {\bibinfo
  {journal} {arXiv:2212.09643}\ } (\bibinfo {year} {2022})}\BibitemShut
  {NoStop}%
\bibitem [{\citenamefont {Giordani}\ \emph {et~al.}(2023)\citenamefont
  {Giordani}, \citenamefont {Mannucci}, \citenamefont {Spagnolo}, \citenamefont
  {Fumero}, \citenamefont {Rampini}, \citenamefont {Rodol{\`a}},\ and\
  \citenamefont {Sciarrino}}]{giordaniCertificationGaussianBoson2023}%
  \BibitemOpen
  \bibfield  {author} {\bibinfo {author} {\bibfnamefont {T.}~\bibnamefont
  {Giordani}}, \bibinfo {author} {\bibfnamefont {V.}~\bibnamefont {Mannucci}},
  \bibinfo {author} {\bibfnamefont {N.}~\bibnamefont {Spagnolo}}, \bibinfo
  {author} {\bibfnamefont {M.}~\bibnamefont {Fumero}}, \bibinfo {author}
  {\bibfnamefont {A.}~\bibnamefont {Rampini}}, \bibinfo {author} {\bibfnamefont
  {E.}~\bibnamefont {Rodol{\`a}}},\ and\ \bibinfo {author} {\bibfnamefont
  {F.}~\bibnamefont {Sciarrino}},\ }\href
  {https://doi.org/10.1088/2058-9565/ac969b} {\bibfield  {journal} {\bibinfo
  {journal} {Quantum Science and Technology}\ }\textbf {\bibinfo {volume}
  {8}},\ \bibinfo {pages} {015005} (\bibinfo {year} {2023})}\BibitemShut
  {NoStop}%
\bibitem [{\citenamefont {Drummond}\ \emph {et~al.}(2022)\citenamefont
  {Drummond}, \citenamefont {Opanchuk}, \citenamefont {Dellios},\ and\
  \citenamefont {Reid}}]{drummondSimulatingComplexNetworks2021}%
  \BibitemOpen
  \bibfield  {author} {\bibinfo {author} {\bibfnamefont {P.~D.}\ \bibnamefont
  {Drummond}}, \bibinfo {author} {\bibfnamefont {B.}~\bibnamefont {Opanchuk}},
  \bibinfo {author} {\bibfnamefont {A.}~\bibnamefont {Dellios}},\ and\ \bibinfo
  {author} {\bibfnamefont {M.~D.}\ \bibnamefont {Reid}},\ }\href
  {https://doi.org/10.1103/PhysRevA.105.012427} {\bibfield  {journal} {\bibinfo
   {journal} {Phys. Rev. A}\ }\textbf {\bibinfo {volume} {105}},\ \bibinfo
  {pages} {012427} (\bibinfo {year} {2022})}\BibitemShut {NoStop}%
\bibitem [{\citenamefont {Popova}\ and\ \citenamefont
  {Rubtsov}(2021)}]{popovaCrackingQuantumAdvantage2021}%
  \BibitemOpen
  \bibfield  {author} {\bibinfo {author} {\bibfnamefont {A.~S.}\ \bibnamefont
  {Popova}}\ and\ \bibinfo {author} {\bibfnamefont {A.~N.}\ \bibnamefont
  {Rubtsov}},\ }\href {http://arxiv.org/abs/2106.01445} {\bibfield  {journal}
  {\bibinfo  {journal} {arXiv:2106.01445}\ } (\bibinfo {year}
  {2021})}\BibitemShut {NoStop}%
\bibitem [{\citenamefont {Grier}\ \emph {et~al.}(2022)\citenamefont {Grier},
  \citenamefont {Brod}, \citenamefont {Arrazola}, \citenamefont {Alonso},\ and\
  \citenamefont {Quesada}}]{grierComplexityBipartiteGaussian2022}%
  \BibitemOpen
  \bibfield  {author} {\bibinfo {author} {\bibfnamefont {D.}~\bibnamefont
  {Grier}}, \bibinfo {author} {\bibfnamefont {D.~J.}\ \bibnamefont {Brod}},
  \bibinfo {author} {\bibfnamefont {J.~M.}\ \bibnamefont {Arrazola}}, \bibinfo
  {author} {\bibfnamefont {M.~B. d.~A.}\ \bibnamefont {Alonso}},\ and\ \bibinfo
  {author} {\bibfnamefont {N.}~\bibnamefont {Quesada}},\ }\href
  {https://doi.org/10.22331/q-2022-11-28-863} {\bibfield  {journal} {\bibinfo
  {journal} {Quantum}\ }\textbf {\bibinfo {volume} {6}},\ \bibinfo {pages}
  {863} (\bibinfo {year} {2022})}\BibitemShut {NoStop}%
\bibitem [{\citenamefont {Lim}\ and\ \citenamefont
  {Oh}(2022)}]{limApproximatingOutcomeProbabilities2022}%
  \BibitemOpen
  \bibfield  {author} {\bibinfo {author} {\bibfnamefont {Y.}~\bibnamefont
  {Lim}}\ and\ \bibinfo {author} {\bibfnamefont {C.}~\bibnamefont {Oh}},\
  }\href {http://arxiv.org/abs/2211.07184} {\bibfield  {journal} {\bibinfo
  {journal} {arXiv:2211.07184}\ } (\bibinfo {year} {2022})}\BibitemShut
  {NoStop}%
\bibitem [{\citenamefont {Dellios}\ \emph
  {et~al.}(2022{\natexlab{b}})\citenamefont {Dellios}, \citenamefont
  {Drummond}, \citenamefont {Opanchuk}, \citenamefont {Teh},\ and\
  \citenamefont {Reid}}]{delliosSimulatingMacroscopicQuantum2022}%
  \BibitemOpen
  \bibfield  {author} {\bibinfo {author} {\bibfnamefont {A.}~\bibnamefont
  {Dellios}}, \bibinfo {author} {\bibfnamefont {P.~D.}\ \bibnamefont
  {Drummond}}, \bibinfo {author} {\bibfnamefont {B.}~\bibnamefont {Opanchuk}},
  \bibinfo {author} {\bibfnamefont {R.~Y.}\ \bibnamefont {Teh}},\ and\ \bibinfo
  {author} {\bibfnamefont {M.~D.}\ \bibnamefont {Reid}},\ }\href
  {https://doi.org/10.1016/j.physleta.2021.127911} {\bibfield  {journal}
  {\bibinfo  {journal} {Physics Letters A}\ }\textbf {\bibinfo {volume}
  {429}},\ \bibinfo {pages} {127911} (\bibinfo {year}
  {2022}{\natexlab{b}})}\BibitemShut {NoStop}%
\bibitem [{\citenamefont {Iosue}\ \emph {et~al.}(2022)\citenamefont {Iosue},
  \citenamefont {Ehrenberg}, \citenamefont {Hangleiter}, \citenamefont
  {Deshpande},\ and\ \citenamefont {Gorshkov}}]{iosue2022page}%
  \BibitemOpen
  \bibfield  {author} {\bibinfo {author} {\bibfnamefont {J.~T.}\ \bibnamefont
  {Iosue}}, \bibinfo {author} {\bibfnamefont {A.}~\bibnamefont {Ehrenberg}},
  \bibinfo {author} {\bibfnamefont {D.}~\bibnamefont {Hangleiter}}, \bibinfo
  {author} {\bibfnamefont {A.}~\bibnamefont {Deshpande}},\ and\ \bibinfo
  {author} {\bibfnamefont {A.~V.}\ \bibnamefont {Gorshkov}},\ }\href
  {https://doi.org/10.48550/arXiv.2209.06838} {\bibfield  {journal} {\bibinfo
  {journal} {arXiv:2209.06838}\ } (\bibinfo {year} {2022})}\BibitemShut
  {NoStop}%
\bibitem [{\citenamefont {Qiao}\ \emph {et~al.}(2022)\citenamefont {Qiao},
  \citenamefont {Huh},\ and\ \citenamefont {Grossmann}}]{qiao2022entanglement}%
  \BibitemOpen
  \bibfield  {author} {\bibinfo {author} {\bibfnamefont {Y.}~\bibnamefont
  {Qiao}}, \bibinfo {author} {\bibfnamefont {J.}~\bibnamefont {Huh}},\ and\
  \bibinfo {author} {\bibfnamefont {F.}~\bibnamefont {Grossmann}},\ }\href
  {https://doi.org/10.48550/arXiv.2210.09915} {\bibfield  {journal} {\bibinfo
  {journal} {arXiv:2210.09915}\ } (\bibinfo {year} {2022})}\BibitemShut
  {NoStop}%
\bibitem [{\citenamefont {Liu}\ \emph {et~al.}(2023)\citenamefont {Liu},
  \citenamefont {Oh}, \citenamefont {Liu}, \citenamefont {Jiang},\ and\
  \citenamefont {Alexeev}}]{liuComplexityGaussianBoson2023}%
  \BibitemOpen
  \bibfield  {author} {\bibinfo {author} {\bibfnamefont {M.}~\bibnamefont
  {Liu}}, \bibinfo {author} {\bibfnamefont {C.}~\bibnamefont {Oh}}, \bibinfo
  {author} {\bibfnamefont {J.}~\bibnamefont {Liu}}, \bibinfo {author}
  {\bibfnamefont {L.}~\bibnamefont {Jiang}},\ and\ \bibinfo {author}
  {\bibfnamefont {Y.}~\bibnamefont {Alexeev}},\ }\href
  {http://arxiv.org/abs/2301.12814} {\bibinfo {title} {Complexity of
  {{Gaussian}} boson sampling with tensor networks}} (\bibinfo {year}
  {2023})\BibitemShut {NoStop}%
\bibitem [{\citenamefont {Oh}\ \emph {et~al.}(2023{\natexlab{a}})\citenamefont
  {Oh}, \citenamefont {Jiang},\ and\ \citenamefont
  {Fefferman}}]{ohClassicalSimulationAlgorithms2023}%
  \BibitemOpen
  \bibfield  {author} {\bibinfo {author} {\bibfnamefont {C.}~\bibnamefont
  {Oh}}, \bibinfo {author} {\bibfnamefont {L.}~\bibnamefont {Jiang}},\ and\
  \bibinfo {author} {\bibfnamefont {B.}~\bibnamefont {Fefferman}},\ }\href
  {http://arxiv.org/abs/2301.11532} {\bibfield  {journal} {\bibinfo  {journal}
  {arXiv:2301.11532}\ } (\bibinfo {year} {2023}{\natexlab{a}})}\BibitemShut
  {NoStop}%
\bibitem [{\citenamefont {{Rahimi-Keshari}}\ \emph {et~al.}(2016)\citenamefont
  {{Rahimi-Keshari}}, \citenamefont {Ralph},\ and\ \citenamefont
  {Caves}}]{rahimi-keshariSufficientConditionsEfficient2016}%
  \BibitemOpen
  \bibfield  {author} {\bibinfo {author} {\bibfnamefont {S.}~\bibnamefont
  {{Rahimi-Keshari}}}, \bibinfo {author} {\bibfnamefont {T.~C.}\ \bibnamefont
  {Ralph}},\ and\ \bibinfo {author} {\bibfnamefont {C.~M.}\ \bibnamefont
  {Caves}},\ }\href {https://doi.org/10.1103/PhysRevX.6.021039} {\bibfield
  {journal} {\bibinfo  {journal} {Physical Review X}\ }\textbf {\bibinfo
  {volume} {6}},\ \bibinfo {pages} {021039} (\bibinfo {year}
  {2016})}\BibitemShut {NoStop}%
\bibitem [{\citenamefont {Deshpande}\ \emph {et~al.}(2018)\citenamefont
  {Deshpande}, \citenamefont {Fefferman}, \citenamefont {Tran}, \citenamefont
  {{Foss-Feig}},\ and\ \citenamefont
  {Gorshkov}}]{deshpandeDynamicalPhaseTransitions2018}%
  \BibitemOpen
  \bibfield  {author} {\bibinfo {author} {\bibfnamefont {A.}~\bibnamefont
  {Deshpande}}, \bibinfo {author} {\bibfnamefont {B.}~\bibnamefont
  {Fefferman}}, \bibinfo {author} {\bibfnamefont {M.~C.}\ \bibnamefont {Tran}},
  \bibinfo {author} {\bibfnamefont {M.}~\bibnamefont {{Foss-Feig}}},\ and\
  \bibinfo {author} {\bibfnamefont {A.~V.}\ \bibnamefont {Gorshkov}},\ }\href
  {https://doi.org/10.1103/PhysRevLett.121.030501} {\bibfield  {journal}
  {\bibinfo  {journal} {Physical Review Letters}\ }\textbf {\bibinfo {volume}
  {121}},\ \bibinfo {pages} {030501} (\bibinfo {year} {2018})}\BibitemShut
  {NoStop}%
\bibitem [{\citenamefont {Chabaud}\ and\ \citenamefont
  {Walschaers}(2023)}]{chabaud2023resources}%
  \BibitemOpen
  \bibfield  {author} {\bibinfo {author} {\bibfnamefont {U.}~\bibnamefont
  {Chabaud}}\ and\ \bibinfo {author} {\bibfnamefont {M.}~\bibnamefont
  {Walschaers}},\ }\href {https://doi.org/10.1103/PhysRevLett.130.090602}
  {\bibfield  {journal} {\bibinfo  {journal} {Physical Review Letters}\
  }\textbf {\bibinfo {volume} {130}},\ \bibinfo {pages} {090602} (\bibinfo
  {year} {2023})}\BibitemShut {NoStop}%
\bibitem [{\citenamefont {Hamilton}\ \emph {et~al.}(2017)\citenamefont
  {Hamilton}, \citenamefont {Kruse}, \citenamefont {Sansoni}, \citenamefont
  {Barkhofen}, \citenamefont {Silberhorn},\ and\ \citenamefont
  {Jex}}]{Hamilton2017}%
  \BibitemOpen
  \bibfield  {author} {\bibinfo {author} {\bibfnamefont {C.~S.}\ \bibnamefont
  {Hamilton}}, \bibinfo {author} {\bibfnamefont {R.}~\bibnamefont {Kruse}},
  \bibinfo {author} {\bibfnamefont {L.}~\bibnamefont {Sansoni}}, \bibinfo
  {author} {\bibfnamefont {S.}~\bibnamefont {Barkhofen}}, \bibinfo {author}
  {\bibfnamefont {C.}~\bibnamefont {Silberhorn}},\ and\ \bibinfo {author}
  {\bibfnamefont {I.}~\bibnamefont {Jex}},\ }\href
  {https://doi.org/10.1103/PhysRevLett.119.170501} {\bibfield  {journal}
  {\bibinfo  {journal} {Phys. Rev. Lett.}\ }\textbf {\bibinfo {volume} {119}},\
  \bibinfo {pages} {170501} (\bibinfo {year} {2017})}\BibitemShut {NoStop}%
\bibitem [{\citenamefont {Quesada}\ \emph {et~al.}(2018)\citenamefont
  {Quesada}, \citenamefont {Arrazola},\ and\ \citenamefont
  {Killoran}}]{quesadaGaussianBosonSampling2018}%
  \BibitemOpen
  \bibfield  {author} {\bibinfo {author} {\bibfnamefont {N.}~\bibnamefont
  {Quesada}}, \bibinfo {author} {\bibfnamefont {J.~M.}\ \bibnamefont
  {Arrazola}},\ and\ \bibinfo {author} {\bibfnamefont {N.}~\bibnamefont
  {Killoran}},\ }\href {https://doi.org/10.1103/PhysRevA.98.062322} {\bibfield
  {journal} {\bibinfo  {journal} {Phys. Rev. A}\ }\textbf {\bibinfo {volume}
  {98}},\ \bibinfo {pages} {062322} (\bibinfo {year} {2018})}\BibitemShut
  {NoStop}%
\bibitem [{\citenamefont {He}\ \emph {et~al.}(2017)\citenamefont {He},
  \citenamefont {Ding}, \citenamefont {Su}, \citenamefont {Huang},
  \citenamefont {Qin}, \citenamefont {Wang} \emph
  {et~al.}}]{heTimeBinEncodedBosonSampling2017}%
  \BibitemOpen
  \bibfield  {author} {\bibinfo {author} {\bibfnamefont {Y.}~\bibnamefont
  {He}}, \bibinfo {author} {\bibfnamefont {X.}~\bibnamefont {Ding}}, \bibinfo
  {author} {\bibfnamefont {Z.-E.}\ \bibnamefont {Su}}, \bibinfo {author}
  {\bibfnamefont {H.-L.}\ \bibnamefont {Huang}}, \bibinfo {author}
  {\bibfnamefont {J.}~\bibnamefont {Qin}}, \bibinfo {author} {\bibfnamefont
  {C.}~\bibnamefont {Wang}}, \emph {et~al.},\ }\href
  {https://doi.org/10.1103/PhysRevLett.118.190501} {\bibfield  {journal}
  {\bibinfo  {journal} {Physical Review Letters}\ }\textbf {\bibinfo {volume}
  {118}},\ \bibinfo {pages} {190501} (\bibinfo {year} {2017})}\BibitemShut
  {NoStop}%
\bibitem [{\citenamefont {Lundeen}\ \emph {et~al.}(2009)\citenamefont
  {Lundeen}, \citenamefont {Feito}, \citenamefont {{Coldenstrodt-Ronge}},
  \citenamefont {Pregnell}, \citenamefont {Silberhorn}, \citenamefont {Ralph},
  \citenamefont {Eisert}, \citenamefont {Plenio},\ and\ \citenamefont
  {Walmsley}}]{lundeenTomographyQuantumDetectors2009}%
  \BibitemOpen
  \bibfield  {author} {\bibinfo {author} {\bibfnamefont {J.~S.}\ \bibnamefont
  {Lundeen}}, \bibinfo {author} {\bibfnamefont {A.}~\bibnamefont {Feito}},
  \bibinfo {author} {\bibfnamefont {H.}~\bibnamefont {{Coldenstrodt-Ronge}}},
  \bibinfo {author} {\bibfnamefont {K.~L.}\ \bibnamefont {Pregnell}}, \bibinfo
  {author} {\bibfnamefont {{\relax Ch}.}~\bibnamefont {Silberhorn}}, \bibinfo
  {author} {\bibfnamefont {T.~C.}\ \bibnamefont {Ralph}}, \bibinfo {author}
  {\bibfnamefont {J.}~\bibnamefont {Eisert}}, \bibinfo {author} {\bibfnamefont
  {M.~B.}\ \bibnamefont {Plenio}},\ and\ \bibinfo {author} {\bibfnamefont
  {I.~A.}\ \bibnamefont {Walmsley}},\ }\href
  {https://doi.org/10.1038/nphys1133} {\bibfield  {journal} {\bibinfo
  {journal} {Nature Physics}\ }\textbf {\bibinfo {volume} {5}},\ \bibinfo
  {pages} {27} (\bibinfo {year} {2009})}\BibitemShut {NoStop}%
\bibitem [{sup()}]{suppl}%
  \BibitemOpen
  \href@noop {} {}\bibinfo {note} {See Supplemental Materials for detailed
  description of 1. Detector tomography of the pseudo-photon-number re- solving
  detectors, 2. Measurement for the matrix phase, 3. Partial photon
  indistinguishability model, 4. Bayesian test results against weaker classical
  hypotheses, 5. On HOG test with the treewidth mockup, which includes Ref.
  \cite{akhlaghi2011nonlinearity,christ2011probing}}\BibitemShut {NoStop}%
\bibitem [{\citenamefont {{Mart{\'i}nez-Cifuentes}}\ \emph
  {et~al.}(2022)\citenamefont {{Mart{\'i}nez-Cifuentes}}, \citenamefont
  {{Fonseca-Romero}},\ and\ \citenamefont {Quesada}}]{ClassicalModelsAre2022}%
  \BibitemOpen
  \bibfield  {author} {\bibinfo {author} {\bibfnamefont {J.}~\bibnamefont
  {{Mart{\'i}nez-Cifuentes}}}, \bibinfo {author} {\bibfnamefont {K.~M.}\
  \bibnamefont {{Fonseca-Romero}}},\ and\ \bibinfo {author} {\bibfnamefont
  {N.}~\bibnamefont {Quesada}},\ }\href {http://arxiv.org/abs/2207.10058}
  {\bibfield  {journal} {\bibinfo  {journal} {arXiv:2207.10058}\ } (\bibinfo
  {year} {2022})}\BibitemShut {NoStop}%
\bibitem [{\citenamefont {Bentivegna}\ \emph {et~al.}(2014)\citenamefont
  {Bentivegna}, \citenamefont {Spagnolo}, \citenamefont {Vitelli},
  \citenamefont {Brod}, \citenamefont {Crespi} \emph
  {et~al.}}]{bentivegnaBayesianApproachBoson2015}%
  \BibitemOpen
  \bibfield  {author} {\bibinfo {author} {\bibfnamefont {M.}~\bibnamefont
  {Bentivegna}}, \bibinfo {author} {\bibfnamefont {N.}~\bibnamefont
  {Spagnolo}}, \bibinfo {author} {\bibfnamefont {C.}~\bibnamefont {Vitelli}},
  \bibinfo {author} {\bibfnamefont {D.~J.}\ \bibnamefont {Brod}}, \bibinfo
  {author} {\bibfnamefont {A.}~\bibnamefont {Crespi}}, \emph {et~al.},\ }\href
  {https://doi.org/10.1142/S021974991560028X} {\bibfield  {journal} {\bibinfo
  {journal} {International Journal of Quantum Information}\ }\textbf {\bibinfo
  {volume} {12}},\ \bibinfo {pages} {1560028} (\bibinfo {year}
  {2014})}\BibitemShut {NoStop}%
\bibitem [{\citenamefont {Barak}\ \emph {et~al.}(2020)\citenamefont {Barak},
  \citenamefont {Chou},\ and\ \citenamefont {Gao}}]{barak2020spoofing}%
  \BibitemOpen
  \bibfield  {author} {\bibinfo {author} {\bibfnamefont {B.}~\bibnamefont
  {Barak}}, \bibinfo {author} {\bibfnamefont {C.-N.}\ \bibnamefont {Chou}},\
  and\ \bibinfo {author} {\bibfnamefont {X.}~\bibnamefont {Gao}},\ }\href@noop
  {} {\bibfield  {journal} {\bibinfo  {journal} {arXiv preprint
  arXiv:2005.02421}\ } (\bibinfo {year} {2020})}\BibitemShut {NoStop}%
\bibitem [{\citenamefont {Gao}\ \emph {et~al.}(2021)\citenamefont {Gao},
  \citenamefont {Kalinowski}, \citenamefont {Chou}, \citenamefont {Lukin},
  \citenamefont {Barak},\ and\ \citenamefont
  {Choi}}]{gaoLimitationsLinearCrossEntropy2021}%
  \BibitemOpen
  \bibfield  {author} {\bibinfo {author} {\bibfnamefont {X.}~\bibnamefont
  {Gao}}, \bibinfo {author} {\bibfnamefont {M.}~\bibnamefont {Kalinowski}},
  \bibinfo {author} {\bibfnamefont {C.-N.}\ \bibnamefont {Chou}}, \bibinfo
  {author} {\bibfnamefont {M.~D.}\ \bibnamefont {Lukin}}, \bibinfo {author}
  {\bibfnamefont {B.}~\bibnamefont {Barak}},\ and\ \bibinfo {author}
  {\bibfnamefont {S.}~\bibnamefont {Choi}},\ }\href
  {http://arxiv.org/abs/2112.01657} {\bibfield  {journal} {\bibinfo  {journal}
  {arXiv:2112.01657}\ } (\bibinfo {year} {2021})}\BibitemShut {NoStop}%
\bibitem [{oh2()}]{oh2023newsim}%
  \BibitemOpen
  \href@noop {} {}\bibinfo {note} {After submission, we became aware of a newly
  developed faster classical approximate simulation algorithm which exhibits
  error level of the HOG score and up-to 5-order correlation function
  comparable to the experiment, by taking advantage of the photon loss
  \cite{Oh2023tensor}. Ref. \cite{Oh2023tensor} takes $\sim 10$ minutes to
  generate the tensor network state corresponding to a specific experiment,
  while in our experiment it takes \SI{1.27}{\us} to generate a sample and
  $\sim 0.1$ ms to reconfigure a new experiment (in \Jiuzhang 2.0, only limited
  by the speed of the physical elements). In addition, Ref. \cite{Oh2023tensor}
  used a simplified model of ground truth for benchmarking of this work,
  without taking pseudo-photon detection and partial distinguishability into
  account. We expect future GBS experiments with higher efficiency, as well as
  better characterization of the experimental imperfections, in particular the
  calibration noise, could also help enhance the quantum advantage against the
  approximate simulation algorithms.}\BibitemShut {Stop}%
\bibitem [{JZ3()}]{JZ3data}%
  \BibitemOpen
  \href {https://quantum.ustc.edu.cn/web/node/1121} {\bibinfo {title} {Raw data
  of {J}iuzhang 3.0}}\BibitemShut {NoStop}%
\bibitem [{\citenamefont {Bressanini}\ \emph {et~al.}(2023)\citenamefont
  {Bressanini}, \citenamefont {Kwon},\ and\ \citenamefont
  {Kim}}]{bressanini2023PPNRDgaussian}%
  \BibitemOpen
  \bibfield  {author} {\bibinfo {author} {\bibfnamefont {G.}~\bibnamefont
  {Bressanini}}, \bibinfo {author} {\bibfnamefont {H.}~\bibnamefont {Kwon}},\
  and\ \bibinfo {author} {\bibfnamefont {M.}~\bibnamefont {Kim}},\ }\href@noop
  {} {\bibfield  {journal} {\bibinfo  {journal} {arXiv preprint
  arXiv:2305.00853}\ } (\bibinfo {year} {2023})}\BibitemShut {NoStop}%
\bibitem [{\citenamefont {Akhlaghi}\ \emph {et~al.}(2011)\citenamefont
  {Akhlaghi}, \citenamefont {Majedi},\ and\ \citenamefont
  {Lundeen}}]{akhlaghi2011nonlinearity}%
  \BibitemOpen
  \bibfield  {author} {\bibinfo {author} {\bibfnamefont {M.~K.}\ \bibnamefont
  {Akhlaghi}}, \bibinfo {author} {\bibfnamefont {A.~H.}\ \bibnamefont
  {Majedi}},\ and\ \bibinfo {author} {\bibfnamefont {J.~S.}\ \bibnamefont
  {Lundeen}},\ }\href@noop {} {\bibfield  {journal} {\bibinfo  {journal}
  {Optics express}\ }\textbf {\bibinfo {volume} {19}},\ \bibinfo {pages}
  {21305} (\bibinfo {year} {2011})}\BibitemShut {NoStop}%
\bibitem [{\citenamefont {Christ}\ \emph {et~al.}(2011)\citenamefont {Christ},
  \citenamefont {Laiho}, \citenamefont {Eckstein}, \citenamefont {Cassemiro},\
  and\ \citenamefont {Silberhorn}}]{christ2011probing}%
  \BibitemOpen
  \bibfield  {author} {\bibinfo {author} {\bibfnamefont {A.}~\bibnamefont
  {Christ}}, \bibinfo {author} {\bibfnamefont {K.}~\bibnamefont {Laiho}},
  \bibinfo {author} {\bibfnamefont {A.}~\bibnamefont {Eckstein}}, \bibinfo
  {author} {\bibfnamefont {K.~N.}\ \bibnamefont {Cassemiro}},\ and\ \bibinfo
  {author} {\bibfnamefont {C.}~\bibnamefont {Silberhorn}},\ }\href@noop {}
  {\bibfield  {journal} {\bibinfo  {journal} {New Journal of Physics}\ }\textbf
  {\bibinfo {volume} {13}},\ \bibinfo {pages} {033027} (\bibinfo {year}
  {2011})}\BibitemShut {NoStop}%
\bibitem [{\citenamefont {Oh}\ \emph {et~al.}(2023{\natexlab{b}})\citenamefont
  {Oh}, \citenamefont {Liu}, \citenamefont {Alexeev}, \citenamefont
  {Fefferman},\ and\ \citenamefont {Jiang}}]{Oh2023tensor}%
  \BibitemOpen
  \bibfield  {author} {\bibinfo {author} {\bibfnamefont {C.}~\bibnamefont
  {Oh}}, \bibinfo {author} {\bibfnamefont {M.}~\bibnamefont {Liu}}, \bibinfo
  {author} {\bibfnamefont {Y.}~\bibnamefont {Alexeev}}, \bibinfo {author}
  {\bibfnamefont {B.}~\bibnamefont {Fefferman}},\ and\ \bibinfo {author}
  {\bibfnamefont {L.}~\bibnamefont {Jiang}},\ }\href@noop {} {\bibfield
  {journal} {\bibinfo  {journal} {arXiv preprint arXiv:2306.03709}\ } (\bibinfo
  {year} {2023}{\natexlab{b}})}\BibitemShut {NoStop}%
\end{thebibliography}
%

\end{document}